\documentclass[12pt]{iopart}

\usepackage{iopams}  
\usepackage{cite}
\usepackage{cleveref}

\newtheorem{proposition}{Proposition}
\newtheorem{corollary}{Corollary}

\renewcommand{\theequation}{\thesection.\arabic{equation}}

\begin{document}

\title[On geometric aspects of the SUSY Fokas--Gel'fand immersion formula]{On geometric aspects of the supersymmetric Fokas--Gel'fand immersion formula}

\author{S. Bertrand}

\address{Department of Mathematics and Statistics, Universit\'e de Montr\'eal,\\ Montr\'eal CP 6128 Succ. Centre-Ville (QC) H3C 3J7, Canada}
\ead{bertrans@crm.umontreal.ca}
\vspace{10pt}

\begin{abstract}
In this paper, we develop a new geometric characterization for the supersymmetric versions of the Fokas--Gel'fand formula for the immersion of soliton supermanifolds with two bosonic and two fermionic independent variables into Lie superalgebras. In order to do so, from a linear spectral problem of a supersymmetric integrable system using the covariant fermionic derivative, we provide a technique to obtain two additional linear spectral problems for that integrable system, one using the bosonic variable derivatives and the other using the fermionic variable derivatives. This allows us to investigate, through the first and second fundamental forms, the geometry of the ($1+1\vert2$)-supermanifolds immersed in Lie superalgebras. Whenever possible, the mean and Gaussian curvatures of the supermanifolds are calculated. These theoretical considerations are applied to the supersymmetric sine-Gordon equation.
\end{abstract}


\vspace{2pc}


\maketitle

\section{Introduction}\label{SecIntro}\setcounter{equation}{0}
The Fokas--Gel'fand formula for the immersion of surfaces into Lie algebras has been at the intersection of several branches of mathematics, such as differential geometry, Lie groups and Lie algebras, complex variables, global analysis, mathematical physics, integrable systems and spectral theory. The Fokas--Gel'fand immersion formula (FGIF) provides rich classes of geometric objects (see e.g. \cite{Sym82,FG96,FGFL,GP11,Cieslinski07,DS92,GLM16,Sym83,Tafel,Cies97}). The FGIF starts from an integrable system of nonlinear partial differential equations with two independent variables, which admits a linear spectral problem (LSP). The compatibility condition of this LSP is assumed to be equivalent to the integrable system for any values of the spectral parameter. One way to introduce the spectral parameter in the LSP is described in \cite{LST,LS90,CGS94} and this method was also extended to supersymmetric (SUSY) integrable systems in \cite{BGH152}. Then, the FGIF considers a first-order infinitesimal deformation of the potential matrices and of the wavefunction, which leaves the LSP and its zero-curvature condition (ZCC) invariant. From these deformations, it is possible to construct an immersion function, its tangent vectors and its normal vector, all taking values in a Lie algebra, under some differential constraint. For a detailed exposition of the theory, see e.g. \cite{Sym82,FG96,FGFL,GP11,Cieslinski07,DS92,GLM16,Sym83,Tafel,Cies97} and references therein. A first deformation was provided by Sym and Tafel \cite{Sym82,Sym83,Tafel}, which considers a deformation of the spectral parameter. Many authors continued to investigate this immersion formula by adding new types of deformations, such as the gauge symmetry deformation (also known as the Cie\'sli\'nski--Doliwa immersion formula) \cite{Cies97,DS92,Cieslinski07} and the (generalized) Lie symmetry approach, see e.g. \cite{FG96,FGFL,Cies97}. The resulting immersion formula is the superposition of all three types deformations. According to \cite{GLM16}, it is possible to link the generalized symmetry approach with the gauge deformation approach. In addition, a geometric characterization of the surface parametrized by the immersion function is provided, which is accomplished using the first and second fundamental forms and the Gaussian and mean curvatures.

The continuous deformations of manifolds are much better understood in the classical case than in their SUSY versions for which only few examples are known. Recently, generalizations of the FGIF were constructed to consider bosonic and fermionic deformations associated with SUSY integrable systems \cite{BG16}. Since soliton supermanifolds associated with SUSY integrable models behave rather differently than the classical smooth manifolds, the rich character of the SUSY versions of the FGIF makes a rather special and interesting object of study. Recent development on SUSY integrable systems, especially on B\"acklund and Darboux transformations, allowed a better understanding and the construction of soliton solutions. For examples, Liu and Ma\~nas \cite{LM98,LM07} provided the $N$-Darboux transformation for the Manin--Radul(--Mathieu) SUSY Korteweg--de Vries (KdV) system, which can also be applied to the SUSY sine-Gordon equaiton, using Pfaffian solutions. In addition, Xue et al. \cite{XLL11} investigated the B\"acklund transformation of the SUSY sine-Gordon equation as a nonlinear superposition formula and they were able to link it with the SUSY MKdV equations. Also, Xue and Liu \cite{XL15} constructed the Darboux and B\"acklund transformations of the SUSY nonlinear Schr\"odinger equations.

The present paper is a follow-up of the investigation performed in \cite{BG16} concerning the SUSY extensions of the FGIF. In the paper \cite{BG16}, the SUSY FGIF only considers (bosonic and fermionic) deformations of the LSP involving covariant fermionic derivatives. Hence, the geometric characterization of the immersion function describes a two-dimensional supermanifold allowing only fermionic infinitesimal displacement in the first and second fundamental forms even though the immersion function depends on two fermionic and two bosonic independent variables. In order to consider the immersion function as a $(1+1\vert2)$-supermanifold, we need to find a way of obtaining more tangent vectors to describe this type of supermanifolds.

Therefore, the aim of this paper is two-fold. The first aim is to construct two additional LSPs, one involving the bosonic independent variable derivatives and the other one involving the fermionic independent variable derivatives. We will use the properties of the fermionic covariant derivatives to construct such LSPs. The second aim is to develop a new geometric characterization of deformation supermanifolds associated with SUSY integrable systems in two bosonic and two fermionic independent variables using the additional LSPs. Hence, we extend the (bosonic and fermionic) deformations to the two additional LSPs, which were not considered in the previous work \cite{BG16}. This characterization is achieved through the construction of structural equations associated with these supermanifolds. We determine their first and second fundamental forms. We find the explicit expressions, whenever possible, for the mean and Gaussian curvatures, which are expressed solely in terms of the potential matrices of the LSP and a normal unit vector. To perform this analysis, the inner product used is the nondegenerate super Killing form on Lie superalgebras based on the supertrace. These theoretical considerations are illustrated by the SUSY sine-Gordon equation (SSGE) for which we obtain richer classes of geometric objects for different deformations than the ones obtained in \cite{BG16}.

The paper is organized as follows. In section~\ref{SecPropLSP}, we state the assumptions made on the SUSY integrable system under consideration and we provide a link between the three different types of LSPs used in this paper. In section~\ref{SecBFGFI}, we construct the bosonic version of the FGIF through deformation supermatrices and a bosonic deformation parameter. In section~\ref{SecFFGFI}, we construct the fermionic version of the FGIF using a fermionic deformation parameter and some deformation supermatrices. In both sections \ref{SecGeomB} and \ref{SecGeomF}, the geometric characterization of their associated SUSY version of the FGIF is provided. In section~\ref{SecsG}, we apply these theoretical considerations to a SUSY integrable system, namely the SSGE, in order to obtain a geometric characterization of the deformation surfaces associated with different symmetries of the LSPs. Some conclusions and future perspectives are discussed in section~\ref{SecConc}.

\section{Properties of the linear spectral problems of a supersymmetric integrable system}\label{SecPropLSP}\setcounter{equation}{0}
Let us consider a SUSY integrable system of PDEs
\begin{equation}
\Delta[u]=0, \qquad [u]=(x_+,x_-,\theta^+,\theta^-,u^1,...,u^k,\partial_{x_\pm}u^1, \partial_{\theta^\pm}u^1,...)\label{PDE}
\end{equation}
written in terms of the bosonic light-cone coordinates $x_+,x_-$, the fermionic independent variables $\theta^+,\theta^-$, the $\mathbb{G}$-valued dependent variables $u^\alpha$ $(\alpha=1,...,k)$, and some derivatives of $u^\alpha$.

Let us assume that there exists a fermionic linear spectral problem (FLSP)
\begin{equation}
D_\pm\Psi=U_\pm\Psi,\label{FLSP}
\end{equation}
where the wavefunction $\Psi=\Psi([u],\lambda)$ takes value in a subgroup $G$ of the $GL(m\vert n,\mathbb{G})$ Lie supergroup and the potential fermionic supermatrices $U_\pm=U_\pm([u],\lambda)$ take values in the associated Lie superalgebra $\mathfrak{g}$, which is a subalgebra of the $\mathfrak{gl}(m\vert n,\mathbb{G})$ Lie superalgebra. The constant $\lambda$ is the spectral parameter taking its values in a subset of $\mathbb{G}$. The covariant derivatives $D_+,D_-$ are given by
\begin{equation}
D_\pm=D_{\theta^\pm}-i\theta^\pm D_{x_\pm},
\end{equation}
where $D_{\theta^\pm}$ and $D_{x_\pm}$ stand for the total derivatives with respect to $\theta^\pm$ and $x_\pm$, respectively. These covariant fermionic derivatives satisfy the properties
\begin{equation}
D_\pm^2=-iD_{x_\pm},\qquad \lbrace D_+,D_-\rbrace=0,\label{DProperty}
\end{equation}
where $\lbrace\cdot,\cdot\rbrace$ stands for the anticommutator. The compatibility condition of equations (\ref{FLSP}) imposed by $\lbrace D_+,D_-\rbrace\Psi=0$ is given by the relation
\begin{equation}
\Omega\equiv D_+U_-+D_-U_+-\lbrace U_+,U_-\rbrace=0,\label{ZCC}
\end{equation}
called the ZCC, which is assumed to be equivalent to the original system (\ref{PDE}) for any possible value of the spectral parameter $\lambda$.

From the FLSP (\ref{FLSP}) and the property (\ref{DProperty}a), we can construct another  LSP which involves the bosonic light-cone coordinate derivatives $D_{x_\pm}$ instead of the covariant derivatives $D_\pm$. This LSP (which, in what follows, is called the $x_\pm$-LSP) takes the form
\begin{equation}
D_{x_\pm}\Psi=V_\pm\Psi,\label{xLSP}
\end{equation}
where the bosonic potential supermatrices $V_\pm=V_\pm([u],\lambda)$ take values in the Lie superalgebra $\mathfrak{g}$ and can be written in terms of the potential supermatrices $U_\pm$ as
\begin{equation}
V_\pm=i\left(D_\pm U_\pm-U_\pm^2\right).\label{VofU}
\end{equation}
The bosonic potential supermatrices $V_\pm$ satisfy the relation (identical to the classical ZCC)
\begin{equation}
D_{x_+}V_--D_{x_-}V_++[V_-,V_+]=0,\label{xZCC}
\end{equation}
which is the compatibility condition of the $x_\pm$-LSP (\ref{xLSP}). The relation (\ref{xZCC}) is satisfied whenever the set of variables $u^\alpha$ is a solution of the original system (\ref{PDE}) and of the ZCC (\ref{ZCC}). However, one should note that equation (\ref{xZCC}) is not necessarily equivalent to the original system (\ref{PDE}), but it is always equivalent at least to certain differential consequences of the original system (\ref{PDE}), i.e.
\begin{equation}
D_+(\Delta[u])=0,\qquad D_-(\Delta[u])=0,\qquad\mbox{or}\qquad D_+D_-(\Delta[u])=0.
\end{equation}
Explicitly, equation (\ref{xZCC}) is given in terms of the ZCC $\Omega$, from equation (\ref{ZCC}), by
\begin{equation}
\begin{array}{l}
D_+D_-\Omega+\lbrace D_+\Omega,U_-\rbrace-\lbrace D_-\Omega,U_+\rbrace+[D_-U_+,\Omega]\\
\hspace{1.5cm}+U_-\Omega U_+-U_+\Omega U_-+\Omega U_+U_--U_-U_+\Omega=0.
\end{array}
\end{equation}

We can also construct a third LSP, which involves the fermionic derivatives $D_{\theta^\pm}$, from the FLSP (\ref{FLSP}) and the $x_\pm$-LSP (\ref{xLSP}). That LSP takes the form
\begin{equation}
D_{\theta^\pm}\Psi=W_\pm\Psi,\label{qLSP}
\end{equation}
which, in what follows, is called the $\theta^\pm$-LSP. The fermionic supermatrices $W_\pm=W_\pm([u],\lambda)$ take values in the Lie superalgebra $\mathfrak{g}$ and can be written in terms of the potential supermatrices $U_\pm$ and $V_\pm$ as
\begin{equation}
W_\pm=U_\pm+i\theta^\pm V_\pm.\label{WofU}
\end{equation}
The ZCC associated with equation (\ref{qLSP}) is given by
\begin{equation}
D_{\theta^+}W_-+D_{\theta^-}W_+-\lbrace W_+,W_-\rbrace=0\label{qZCC}
\end{equation}
and can be written in terms of the supermatrices $U_\pm$ and $V_\pm$ as
\begin{equation}
\hspace{-2cm}\begin{array}{l}
\left(D_+U_-+D_-U_+-\lbrace U_+,U_-\rbrace\right)+i\theta^+\left(D_{x_+}U_--D_-V_++[U_-,V_+]\right)\\
+i\theta^-\left(D_{x_-}U_+-D_+V_-+[U_+,V_-]\right)+\theta^+\theta^-\left(D_{x_-}V_+-D_{x_+}V_-+[V_+,V_-]\right)=0.
\end{array}\label{qZCCdec}
\end{equation}
The second and third sets of terms in equation (\ref{qZCCdec}) represent the mixed compatibility conditions of the FLSP (\ref{FLSP}) and the $x_\pm$-LSP (\ref{xLSP}), i.e.
\begin{equation}
0=[D_{x_\pm},D_\mp]\Psi=\left(D_{x_\pm}U_\mp-D_\mp V_\pm+[U_\mp,V_\pm]\right)\Psi.
\end{equation}
These compatibility conditions are equivalent to the relation
\begin{equation}
0=D_\pm\Omega+[\Omega,U_\pm],
\end{equation}
which is in the form of a Lax equation \cite{Helein}.

\section{The bosonic supersymmetric Fokas--Gel'fand immersion formula}\label{SecBFGFI}\setcounter{equation}{0}
Let us consider a deformation of the three LSPs (\ref{FLSP}), (\ref{xLSP}) and (\ref{qLSP}) which preserves their forms, i.e.
\begin{equation}
D_\pm\tilde{\Psi}=\tilde{U}_\pm\tilde{\Psi},\qquad D_{x_\pm}\tilde{\Psi}=\tilde{V}_\pm\tilde{\Psi},\qquad D_{\theta^\pm}\tilde{\Psi}=\tilde{W}_\pm\tilde{\Psi},\label{BdLSP}
\end{equation}
for which their compatibility conditions are equivalent to the original system (\ref{PDE}) for any value of the spectral parameter $\lambda$ and such that the deformed potential supermatrices are given by
\begin{equation}
\hspace{-1cm}\tilde{U}_\pm=U_\pm+\epsilon A_\pm,\quad \tilde{V}_\pm=V_\pm+\epsilon V_\pm,\quad \tilde{W}_\pm=W_\pm+\epsilon C_\pm \in\mathfrak{g}\label{Bpotdef}
\end{equation}
together with the deformed wavefunction $\tilde{\Psi}$ written in terms of a deformed supermanifold $F=F([u],\lambda)$ as
\begin{equation}
\tilde{\Psi}=\Psi(I+\epsilon F)\in G.\label{BPsidef}
\end{equation}
The deformation fermionic supermatrices $A_\pm=A_\pm([u],\lambda),C_\pm=C_\pm([u],\lambda)$ and bosonic supermatrices $B_\pm=B_\pm([u],\lambda),F$ take their values in the Lie superalgebra $\mathfrak{g}$ and the bosonic deformation parameter $\epsilon$ is assumed to vanish for quadratic terms, i.e. $\epsilon^2=0$. Hence, the bosonic parameter $\epsilon$ is either very small or nilpotent of order 2. The compatibility conditions of the deformed LSPs (\ref{BdLSP}) impose conditions on the supermatrices $A_\pm$, $B_\pm$ and $C_\pm$ given by the relations
\begin{equation}
\begin{array}{l}
D_+A_-+D_-A_+-\lbrace A_+,U_-\rbrace-\lbrace A_-,U_+\rbrace=0,\\
D_{x_+}B_--D_{x_-}B_++[B_-,V_+]+[V_-,B_+]=0,\\
D_{\theta^+}C_-+D_{\theta^-}C_+-\lbrace C_+,W_-\rbrace-\lbrace C_-,W_+\rbrace=0,
\end{array}\label{BIDZCC}
\end{equation}
which are equivalent to the deformation of the three ZCCs (\ref{ZCC}), (\ref{xZCC}) and (\ref{qZCC}), respectively, up to the addition of supermatrices at least of first order in $\epsilon$.

From equations (\ref{FLSP}), (\ref{BdLSP}a), (\ref{Bpotdef}a) and (\ref{BPsidef}), we can compute the covariant derivatives of the deformed surface $F$, which are given by
\begin{equation}
D_\pm F=\Psi^{-1}A_\pm\Psi\label{BtanA}
\end{equation}
up to the addition of fermionic supermatrices which are at least of first order in $\epsilon$. Similarly, we can compute the tangent vectors of the deformed surfaces $F$ in the $x_\pm$- and $\theta^\pm$-directions, which take the forms
\begin{equation}
D_{x_\pm}F=\Psi^{-1}B_\pm\Psi,\qquad D_{\theta^\pm}F=\Psi^{-1}C_\pm\Psi\label{Bvtan}
\end{equation}
up to the addition of supermatrices which are at least of first order in $\epsilon$.

Furthermore, from equations (\ref{BtanA}), it is possible to determine the supermatrices $B_\pm$ and $C_\pm$ explicitly in terms of $A_\pm$ and the potential supermatrices $U_\pm$, i.e.
\begin{equation}
B_\pm=i\left(D_\pm A_\pm-\lbrace U_\pm,A_\pm\rbrace\right),\qquad C_\pm=A_\pm+i\theta^\pm B_\pm,\label{BBCofA}
\end{equation}
which are equivalent to the deformations of equations (\ref{VofU}) and (\ref{WofU}), respectively.

A SUSY analogue of the classical Fokas--Gel'fand theorem containing the main results on the immersion of surfaces in Lie algebras can be formulated.
The supermanifold $F$ consists of the superposition of three terms, i.e.
\begin{equation}
F=\beta(\lambda)\Psi^{-1}\partial_\lambda\Psi+\Psi^{-1}\mbox{pr}(\omega_R)\Psi+\Psi^{-1}S\Psi.\label{BF}
\end{equation}
The first term refers to the Sym--Tafel immersion formula, which represents a deformation in the spectral parameter $\lambda$ generated by the vector field $\beta(\lambda)\partial_\lambda$, where $\beta(\lambda)$ is a function of $\lambda$ with the constraint $\deg(\beta)=\deg(\lambda)$. The second term represents a Lie symmetry deformation generated by the bosonic vector field $\omega_R$, which is common to both the original system (\ref{PDE}) (and the ZCC (\ref{ZCC})) and the FSLP (\ref{FLSP}). In this paper, we use a SUSY adaptation of the formalism of the prolongation of a vector field, which is described in \cite{Olver93}. The last term represents a left-transformation of the wavefunction $\Psi$ by a Lie supergroup gauge generated by the conjugated supermatrix $S$ taking its value in the bosonic part of the Lie superalgebra $\mathfrak{g}$. The associated supermatrices $A_\pm$, $B_\pm$ and $C_\pm$ take the forms
\begin{equation}
\hspace{-1.5cm}\begin{array}{l}
A_\pm=\beta(\lambda)\partial_\lambda U_\pm+\left(\mbox{pr}(\omega_R)U_\pm+[D_\pm,\mbox{pr}(\omega_R)]\Psi\Psi^{-1}\right)+\left(D_\pm S+[S,U_\pm]\right),\\
B_\pm=\beta(\lambda)\partial_\lambda V_\pm+\left(\mbox{pr}(\omega_R)V_\pm+[D_{x_\pm},\mbox{pr}(\omega_R)]\Psi\Psi^{-1}\right)+\left(D_{x_\pm} S+[S,V_\pm]\right),\\
C_\pm=\beta(\lambda)\partial_\lambda W_\pm+\left(\mbox{pr}(\omega_R)W_\pm+[D_{\theta^\pm},\mbox{pr}(\omega_R)]\Psi\Psi^{-1}\right)+\left(D_{\theta^\pm} S+[S,W_\pm]\right).
\end{array}\label{BABC}
\end{equation}

Under these assumptions, we have the following statements:
\begin{proposition}\label{PropB}
If we consider the bosonic deformations (\ref{Bpotdef}) and (\ref{BPsidef}) which preserve the LSPs (\ref{FLSP}), (\ref{xLSP}) and (\ref{qLSP}) and their ZCCs (i.e. satisfy equations (\ref{BdLSP}) and (\ref{BIDZCC})) where $F$, $B_\pm$ are bosonic supermatrices and $A_\pm$, $C_\pm$ are fermionic supermatrices in the Lie superalgebra $\mathfrak{g}$ and $\epsilon$ is a bosonic parameter such that $\epsilon^2$ vanishes, then there exists an immersion superfield $F$ which defines $(2\vert 1+1)$-dimensional supermanifolds immersed in the Lie superalgebra $\mathfrak{g}$ provided that its tangent vectors (\ref{Bvtan}) are linearly independent.
\end{proposition}

\begin{corollary}\label{CoroB}
If we consider the deformed supermanifold $F$, as defined in Proposition~\ref{PropB}, in the form (\ref{BF}) where $\beta(\lambda)$ is a function of the spectral parameter $\lambda$ with the constraint $\deg(\beta)=\deg(\lambda)$, $\mbox{pr}(\omega_R)$ is the prolongation of a bosonic supervector field $\omega_R$ which is associated with a common (generalized) symmetry of the original system (\ref{PDE}) and the LSPs (\ref{FLSP}), (\ref{xLSP}) and (\ref{qLSP}), then $F$ is a solution of the equations (\ref{BIDZCC}) and (\ref{BtanA}) for which the supermatrices $A_\pm$, $B_\pm$ and $C_\pm$ take the form (\ref{BABC}).
\end{corollary}

\subsection{The geometric characterization associated with the bosonic Fokas--Gel'fand immersion formula}\label{SecGeomB}
Different geometrical approaches and forms can be used depending on the interpretation of the physical model or system under investigation \cite{Cornwell,DeWitt,Freed,Varadarajan}. In this paper, we focus our study on a SUSY version of the Killing form on Lie superalgebras, which is given by the supertrace,
\begin{equation}
\hspace{-1.5cm}\langle M,N\rangle=\frac{1}{2}\mbox{str}(MN)=\frac{1}{2}\tr\left(E^{\deg(MN)+1}MN\right),\quad M,N\in\mathfrak{g}\subset\mathfrak{gl}(m\vert n,\mathbb{G}),
\end{equation}
where
\begin{equation}
E=\left(\begin{array}{cc}
I_m & 0 \\
0 & -I_n
\end{array}\right),
\end{equation}
that provides a pseudo-Riemannian description of the associated geometry whenever it is nondegenerate. The factor $\frac{1}{2}$ has been selected for convenience for the $\mathfrak{sl}(2\vert1,\mathbb{G})$ Lie superalgebra in the example of the SUSY sine-Gordon equation. Throughout the geometric characterization, we use the abbreviated notation for the indices associated with the independent bosonic and fermionic variables $x_\pm$ and $\theta^\pm$ given by
\begin{equation}
\begin{array}{llll}
x_+\rightarrow1, & D_{x_+}=D_1, & \theta^+\rightarrow3, & D_{\theta^+}=D_3,\\
x_-\rightarrow2, & D_{x_-}=D_2, & \theta^-\rightarrow4, & D_{\theta^-}=D_4.
\end{array}
\end{equation}

The metric coefficients associated with the bosonic SUSY FGIF are given by
\begin{equation}
g_{ij}=\langle D_iF,D_jF\rangle,\qquad i,j,=1,2,3,4.
\end{equation}
The 16 coefficients can be written in terms of the supermatrices $B_\pm$ and $C_\pm$ as
\begin{equation}
\hspace{-1cm}\begin{array}{lll}
g_{11}=\langle B_+,B_+\rangle, & g_{12}=g_{21}=\langle B_+,B_-\rangle,& g_{22}=\langle B_-,B_-\rangle,\\
g_{13}=g_{31}=\langle B_+,C_+\rangle, & g_{14}=g_{41}=\langle B_+,C_-\rangle,& g_{33}=g_{44}=0,\\
g_{23}=g_{32}=\langle B_-,C_+\rangle, & g_{24}=g_{42}=\langle B_-,C_-\rangle,& g_{34}=-g_{43}=\langle C_+,C_-\rangle.
\end{array}\label{Bgij}
\end{equation}
The metric coefficients can be represented by a bosonic $\mathfrak{gl}(2\vert2,\mathbb{G})$-valued supermatrix,
\begin{equation}
g\equiv [g_{ij}]=\left(\begin{array}{cc|cc}
g_{11} & g_{12} & g_{13} & g_{14} \\
g_{12} & g_{22} & g_{23} & g_{24} \\
\hline g_{13} & g_{23} & 0 & g_{34} \\
g_{14} & g_{24} & -g_{34} & 0
\end{array}\right),\label{Bg}
\end{equation}
which depends on a maximum of 8 linearly independent coefficients. The first fundamental form is defined to be
\begin{equation}
\hspace{-2.2cm}\begin{array}{l}
I=\sum_{i,j=1}^4g_{ij}d_id_j=g_{11}(dx_+)^2+2g_{12}dx_+dx_-+g_{22}(dx_-)^2+2g_{34}d\theta^+d\theta^-\\
\hspace{4cm}+2\left(g_{13}dx^+d\theta^++g_{14}dx^+d\theta^-+g_{23}dx_-d\theta^++g_{24}dx_-d\theta^-\right).
\end{array}
\end{equation}
Using equations (\ref{BBCofA}), we can eliminate the supermatrices $C_\pm$ in the metric coefficients (\ref{Bgij}) and reduce some coefficients in terms of the supermatrices $A_\pm$, $B_\pm$, which are given by
\begin{equation}
\begin{array}{l}
g_{13}=i\langle D_+A_+,A_+\rangle+i\theta^+ g_{11},\qquad g_{14}=\langle B_+,A_-\rangle+i\theta^- g_{12},\\
g_{23}=\langle B_-,A_+\rangle+i\theta^+ g_{12},\qquad g_{24}=i\langle D_-A_-,A_-\rangle+i\theta^- g_{22},\\
g_{34}=\langle A_+,A_-\rangle+i\theta^+ g_{14}-i\theta^-g_{23}+\theta^+\theta^-g_{12}.
\end{array}
\end{equation}

In order to construct the second fundamental form, we consider a normal unit vector $N$ taking the form of a bosonic $\mathfrak{g}$-valued supermatrix which has the properties
\begin{equation}
\langle N,N\rangle=1,\qquad \langle D_iF,N\rangle=0,\qquad i=1,2,3,4.
\end{equation}
The second fundamental form's coefficients are defined by
\begin{equation}
b_{ij}=\langle D_jD_iF,N\rangle,\qquad i,j=1,2,3,4\label{Bbij}
\end{equation}
and they can be represented in the same form as equation (\ref{Bg}) with a maximum of 8 linearly independent coefficients,
\begin{equation}
b\equiv [b_{ij}]=\left(\begin{array}{cc|cc}
b_{11} & b_{12} & b_{13} & b_{14} \\
b_{12} & b_{22} & b_{23} & b_{24} \\
\hline b_{13} & b_{23} & 0 & b_{34} \\
b_{14} & b_{24} & -b_{34} & 0
\end{array}\right).\label{Bb}
\end{equation}
The second fundamental form then takes the form
\begin{equation}
\begin{array}{l}
II=b_{11}(dx_+)^2+2b_{12}dx_+dx_-+b_{22}(dx_-)^2+2b_{34}d\theta^+d\theta^-\\
\hspace{1cm}+2\left(b_{13}dx^+d\theta^++b_{14}dx^+d\theta^-+b_{23}dx_-d\theta^++b_{24}dx_-d\theta^-\right).
\end{array}
\end{equation}
Eliminating the supermatrices $W_\pm$ and $C_\pm$ in (\ref{Bbij}), we obtain
\begin{equation}
\begin{array}{l}
b_{13}=i\langle D_+^3F,N\rangle+i\theta^+b_{11},\qquad b_{14}=i\langle D_+^2D_-F,N\rangle+i\theta^-b_{12},\\
b_{23}=i\langle D_-D_+^2F,N\rangle+i\theta^+b_{12},\qquad b_{24}=i\langle D_-^3F,N\rangle+i\theta^-b_{22},\\
b_{34}=\langle D_-D_+F,N\rangle-i\theta^+b_{14}+i\theta^-b_{23}-\theta^+\theta^-b_{12}.
\end{array}
\end{equation}
Whenever the bosonic supermatrix $g$ in equation (\ref{Bg}) is invertible, we can compute the mean curvature given by
\begin{equation}
H=\frac{1}{4}\mbox{str}(bg^{-1})
\end{equation}
and if $b$ is also invertible, then the Gaussian curvature can be computed using the superdeterminant,
\begin{equation}
K=\mbox{sdet}(bg^{-1}).
\end{equation}

\section{The fermionic supersymmetric Fokas--Gel'fand immersion formula}\label{SecFFGFI}\setcounter{equation}{0}
Let us consider a deformation of the three LSPs (\ref{FLSP}), (\ref{xLSP}) and (\ref{qLSP}) which leaves these LSPs invariant, i.e.
\begin{equation}
D_\pm\tilde{\Psi}=\tilde{U}_\pm\tilde{\Psi},\qquad D_{x_\pm}\tilde{\Psi}=\tilde{V}_\pm\tilde{\Psi},\qquad D_{\theta^\pm}\tilde{\Psi}=\tilde{W}_\pm\tilde{\Psi},\label{FdLSP}
\end{equation}
and whose compatibility conditions are equivalent to the original system (\ref{PDE}) for any value of the spectral parameter $\lambda$. The deformed potential supermatrices are given by
\begin{equation}
\hspace{-2cm}\tilde{U}_\pm=U_\pm+\epsilon A_\pm,\qquad \tilde{V}_\pm=V_\pm+\epsilon B_\pm,\qquad \tilde{W}_\pm=W_\pm+\epsilon C_\pm,\quad\in\mathfrak{g}\label{Fpotdef}
\end{equation}
and the deformed wavefunction $\tilde{\Psi}$ is written in terms of a deformed supermanifold $F=F([u],\lambda)$,
\begin{equation}
\tilde{\Psi}=\Psi(I+\epsilon F)\in G.\label{FPsidef}
\end{equation}
The bosonic deformation supermatrices $A_\pm=A_\pm([u],\lambda)$, $C_\pm=C_\pm([u],\lambda)$ and the fermionic deformation supermatrices $B_\pm=B_\pm([u],\lambda)$, $F$ take their values in the Lie superalgebra $\mathfrak{g}$ and $\epsilon$ is a fermionic deformation parameter. The compatibility conditions of the deformed LSPs (\ref{FdLSP}) impose conditions on the supermatrices $A_\pm$, $B_\pm$ and $C_\pm$ given by the relations
\begin{equation}
\begin{array}{l}
D_+A_-+D_-A_++[A_+,U_-]+[A_-,U_+]=0,\\
D_{x_+}B_--D_{x_-}B_++[V_-,B_+]+[B_-,V_+]=0,\\
D_{\theta^+}C_-+D_{\theta^-}C_++[C_-,W_+]+[C_+,W_-]=0,
\end{array}\label{FIDZCC}
\end{equation}
up to the addition of supermatrices which are at least of first order in $\epsilon$. Equations (\ref{FIDZCC}) are equivalent to the deformation of the three ZCCs (\ref{ZCC}), (\ref{xZCC}) and (\ref{qZCC}).

From equations (\ref{FLSP}), (\ref{FdLSP}a), (\ref{Fpotdef}a) and (\ref{FPsidef}), one can compute the covariant derivatives of the deformed surface $F$, which is given by
\begin{equation}
D_\pm F=-\Psi^{-1}A_\pm\Psi.\label{FtanA}
\end{equation}
Similarly, we can compute the tangent vectors of the deformed surface $F$ in the $x_\pm$- and $\theta^\pm$-directions, which take the forms
\begin{equation}
D_{x_\pm}F=\Psi^{-1}B_\pm\Psi,\qquad D_{\theta^\pm}F=-\Psi^{-1}C_\pm\Psi\label{Fvtan}
\end{equation}
up to the addition of supermatrices which are at least of first order in $\epsilon$.

Furthermore, from equation (\ref{FtanA}), it is possible to determine the supermatrices $B_\pm$ and $C_\pm$ explicitly in terms of $A_\pm$ and the potential supermatrices $U_\pm$,
\begin{equation}
B_\pm=-i\left(D_\pm A_\pm+[A_\pm,U_\pm]\right),\qquad C_\pm=A_\pm-i\theta^\pm B_\pm,\label{FBCofA}
\end{equation}
which are equivalent to the deformations of equations (\ref{VofU}) and (\ref{WofU}), respectively.

Similarly to the bosonic and classical case, the supermanifold $F$ consists of the superposition of three terms, i.e
\begin{equation}
F=\beta(\lambda)\Psi^{-1}\partial_\lambda\Psi+\Psi^{-1}\mbox{pr}(\omega_R)\Psi+\Psi^{-1}S\Psi.\label{FF}
\end{equation}
The first term refers to the Sym--Tafel immersion formula, which represents a deformation in the spectral parameter $\lambda$ generated by the vector field $\beta(\lambda)\partial_\lambda$, where $\beta(\lambda)$ is a function of $\lambda$ with the constraint $\deg(\beta)\neq\deg(\lambda)$. The second term represents a Lie symmetry deformation generated by the fermionic vector field $\omega_R$, which is common to both the original system (\ref{PDE}) (and the ZCC (\ref{ZCC})) and the FLSP (\ref{FLSP}). The last term represents a left-transformation of the wavefunction $\Psi$ by a Lie supergroup gauge generated by the conjugated supermatrix $S$ taking its values in the fermionic part of the $\mathfrak{gl}(m\vert n,\mathbb{G})$ Lie superalgebra. The associated supermatrices $A_\pm$, $B_\pm$ and $C_\pm$ take the forms
\begin{equation}
\hspace{-2cm}\begin{array}{l}
A_\pm=\beta(\lambda)\partial_\lambda U_\pm+\left(\mbox{pr}(\omega_R)U_\pm-\lbrace D_\pm,\mbox{pr}(\omega_R)\rbrace\Psi\Psi^{-1}\right)+\left(-D_\pm S+\lbrace S,U_\pm\rbrace\right),\\
B_\pm=\beta(\lambda)\partial_\lambda V_\pm+\left(\mbox{pr}(\omega_R)V_\pm-[D_{x_\pm},\mbox{pr}(\omega_R)]\Psi\Psi^{-1}\right)+\left(D_{x_\pm} S+[S,V_\pm]\right),\\
C_\pm=\beta(\lambda)\partial_\lambda W_\pm+\left(\mbox{pr}(\omega_R)W_\pm-\lbrace D_{\theta^\pm},\mbox{pr}(\omega_R)\rbrace\Psi\Psi^{-1}\right)+\left(-D_{\theta^\pm} S+\lbrace S,W_\pm\rbrace\right).
\end{array}\label{FABC}
\end{equation}

Under these assumptions, we have the following statements:
\begin{proposition}\label{PropF}
If we consider the fermionic deformations (\ref{Fpotdef}) and (\ref{FPsidef}) which preserve the LSPs (\ref{FLSP}), (\ref{xLSP}) and (\ref{qLSP}) and their ZCCs, i.e. satisfy equations (\ref{FdLSP}) and (\ref{FIDZCC}), where $F$, $B_\pm$ are fermionic supermatrices and $A_\pm$, $C_\pm$ are bosonic supermatrices taking values in the Lie superalgebra $\mathfrak{g}$ and $\epsilon$ is a fermionic parameter, then there exists an immersion superfield $F$ which defines $(2\vert 1+1)$-dimensional supermanifolds immersed in the Lie superalgebra $\mathfrak{g}$ provided that its tangent vectors (\ref{Fvtan}) are linearly independent.
\end{proposition}

\begin{corollary}\label{CoroF}
If we consider the deformed supermanifold $F$, as defined in Proposition \ref{PropF}, in the form (\ref{FF}), where $\beta(\lambda)$ is a function of the spectral parameter $\lambda$ with the constraint $\deg(\beta)\neq\deg(\lambda)$, $\mbox{pr}(\omega_R)$ is the prolongation of a fermionic vector superfield $\omega_R$ which is associated with a common (generalized) symmetry of the original system (\ref{PDE}) and the LSPs (\ref{FLSP}), (\ref{xLSP}) and (\ref{qLSP}), then $F$ is a solution of equations (\ref{FIDZCC}) and (\ref{FtanA}) such that the supermatrices $A_\pm$, $B_\pm$ and $C_\pm$ take the form (\ref{FABC}).
\end{corollary}

\subsection{The geometric characterization associated with the fermionic Fokas--Gel'fand immersion formula}\label{SecGeomF}
The metric coefficients associated with the fermionic SUSY FGIF are given by
\begin{equation}
g_{ij}=\langle D_iF,D_jF\rangle,\qquad i,j=1,2,3,4.
\end{equation}
The 16 coefficients can be written in terms of the supermatrices $B_\pm$ and $C_\pm$ as
\begin{equation}
\hspace{-2cm}\begin{array}{lll}
g_{11}=g_{22}=0, & g_{12}=-g_{21}=\langle B_+,B_-\rangle, & g_{13}=g_{31}=-\langle B_+,C_+\rangle,\\
g_{14}=g_{41}=-\langle B_+,C_-\rangle, & g_{23}=g_{32}=-\langle B_-,C_+\rangle, & g_{24}=g_{42}=-\langle B_-,C_-\rangle,\\
g_{33}=\langle C_+,C_+\rangle, & g_{34}=g_{43}=\langle C_+,C_-\rangle, & g_{44}=\langle C_-,C_-\rangle.
\end{array}
\end{equation}
The metric coefficients can be represented by a bosonic $\mathfrak{gl}(2\vert2,\mathbb{G})$-valued supermatrix,
\begin{equation}
g\equiv [g_{ij}]=\left(\begin{array}{cc|cc}
0 & g_{12} & g_{13} & g_{14} \\
-g_{12} & 0 & g_{23} & g_{24} \\
\hline g_{13} & g_{23} & g_{33} & g_{34} \\
g_{14} & g_{24} & g_{34} & g_{44}
\end{array}\right),\label{Fg}
\end{equation}
which depends on a maximum of 8 linearly independent coefficients. The first fundamental form is defined to be
\begin{equation}
\hspace{-2cm}I=\sum_{i,j=1}^4g_{ij}d_id_j=2\left(g_{13}dx_+d\theta^++g_{14}dx_+d\theta^-+g_{23}dx_-d\theta^++g_{24}dx_-d\theta^-\right).
\end{equation}
Using equation (\ref{FBCofA}), we can eliminate the supermatrices $C_\pm$ in some of the coefficients of the metric and reduce them in terms of the supermatrices $A_\pm$ and $B_\pm$, which are given by
\begin{equation}
\hspace{-2.5cm}\begin{array}{l}
g_{13}=-\langle B_+,A_+\rangle,\quad g_{14}=-\langle B_+,A_-\rangle-i\theta^-g_{12},\quad g_{23}=-\langle B_-,A_+\rangle+i\theta^+g_{12},\\
g_{24}=-\langle B_-,A_-\rangle, \quad g_{33}=\langle A_+,A_+\rangle+2i\theta^+g_{13},\quad g_{44}=\langle A_-,A_-\rangle+2i\theta^-g_{24},\\
g_{34}=\langle A_+,A_-\rangle+i\theta^+g_{14}+i\theta^-g_{23}-\theta^+\theta^-g_{12}.
\end{array}
\end{equation}
In order to construct the second fundamental form, we consider a normal unit vector $N$ which takes the form of a bosonic $\mathfrak{g}$-valued supermatrix and has the following properties:
\begin{equation}
\langle N,N\rangle=1,\qquad \langle D_iF,N\rangle=0,\qquad i=1,2,3,4.
\end{equation}
The second fundamental form's coefficients are defined by
\begin{equation}
b_{ij}=\langle D_jD_iF,N\rangle,\qquad i,j=1,2,3,4\label{Fbij}
\end{equation}
and they can be represented in the same form as equation (\ref{Bb}) but as a fermionic supermatrix with a maximum of 8 linearly independent coefficients,
\begin{equation}
b\equiv [b_{ij}]=\left(\begin{array}{cc|cc}
b_{11} & b_{12} & b_{13} & b_{14} \\
b_{12} & b_{22} & b_{23} & b_{24} \\
\hline b_{13} & b_{23} & 0 & b_{34} \\
b_{14} & b_{24} & -b_{34} & 0
\end{array}\right).
\end{equation}
By eliminating the supermatrices $W_\pm$ and $C_\pm$ in (\ref{Fbij}), we obtain, similarly to the bosonic case,
\begin{equation}
\begin{array}{l}
b_{13}=i\langle D_+^3F,N\rangle+i\theta^+b_{11},\qquad b_{14}=i\langle D_+^2D_-F,N\rangle+i\theta^-b_{12},\\
b_{23}=i\langle D_-D_+^2F,N\rangle+i\theta^+b_{12},\qquad b_{24}=i\langle D_-^3F,N\rangle+i\theta^-b_{22},\\
b_{34}=\langle D_-D_+F,N\rangle-i\theta^+b_{14}+i\theta^-b_{23}-\theta^+\theta^-b_{12}.
\end{array}
\end{equation}
Whenever the bosonic supermatrix $g$ is invertible, we can compute the mean curvature,
\begin{equation}
H=\frac{1}{4}\mbox{str}(bg^{-1}).
\end{equation}
However, the Gaussian curvature $K$ is not defined since $b$ is a fermionic supermatrix.

\section{Example: The supersymmetric sine-Gordon equation}\label{SecsG}\setcounter{equation}{0}
Let us consider the SSGE
\begin{equation}
D_+D_-\phi=i\sin\phi,\label{sG}
\end{equation}
where $\phi=\phi(x_+,x_-,\theta^+,\theta^-)$ is a bosonic $\mathbb{G}$-valued function. Its associated FLSP (\ref{FLSP}) is defined by the potential matrices \cite{SHS06,Bertrand16,CK78}
\begin{equation}
\hspace{-2.5cm}U_+=\frac{1}{2\sqrt{\lambda}}\left(\begin{array}{ccc}
0 & 0 & ie^{i\phi} \\
0 & 0 & -ie^{-i\phi} \\
-e^{-i\phi} & e^{i\phi} & 0
\end{array}\right),\qquad U_-=\left(\begin{array}{ccc}
iD_-\phi & 0 & -i\sqrt{\lambda} \\
0 & -iD_-\phi & i\sqrt{\lambda} \\
-\sqrt{\lambda} & \sqrt{\lambda} & 0
\end{array}\right),
\end{equation}
which take their values in the $\mathfrak{sl}(2\vert1,\mathbb{G})$ Lie superalgebra. The associated $x_\pm$-LSP (\ref{xLSP}) is composed of the $\mathfrak{sl}(2\vert1,\mathbb{G})$-valued bosonic supermatrices
\begin{equation}
\begin{array}{l}
\vspace{3mm}V_+=\frac{1}{2}\left(\begin{array}{ccc}
-1/2\lambda & e^{2i\phi}/2\lambda & -iD_+\phi e^{i\phi}/\sqrt{\lambda} \\
e^{-2i\phi}/2\lambda & -1/2\lambda & -iD_+\phi e^{-i\phi}/\sqrt{\lambda} \\
D_+\phi e^{-i\phi}/\sqrt{\lambda} & D_+\phi e^{i\phi}/\sqrt{\lambda} & -1/\lambda
\end{array}\right),\\
V_-=\left(\begin{array}{ccc}
i\partial_{x_-}\phi+\lambda & -\lambda & -i\sqrt{\lambda}D_-\phi \\
-\lambda & -i\partial_{x_-}\phi+\lambda & -i\sqrt{\lambda}D_-\phi \\
-\sqrt{\lambda}D_-\phi & -\sqrt{\lambda}D_-\phi & 2\lambda
\end{array}\right),
\end{array}
\end{equation}
and the $\mathfrak{sl}(2\vert1,\mathbb{G})$-valued fermionic supermatrices $W_\pm$ in the $\theta^\pm$-LSP (\ref{qLSP}) take the forms
\begin{equation*}
\hspace{-2.5cm}\begin{array}{l}
\vspace{3mm}W_+=\frac{1}{2\lambda}\left(\begin{array}{ccc}
-i\theta^+/2 & i\theta^+e^{2i\phi}/2 & i\sqrt{\lambda}e^{i\phi}(1-i\theta^+\partial_{\theta^+}\phi) \\
i\theta^+e^{-2i\phi}/2 & -i\theta^+/2 & -i\sqrt{\lambda}e^{-i\phi}(1+i\theta^+\partial_{\theta^+}\phi) \\
-\sqrt{\lambda}e^{-i\phi}(1+i\theta^+\partial_{\theta^+}\phi) & \sqrt{\lambda}e^{i\phi}(1-i\theta^+\partial_{\theta^+}\phi) & i\theta^+
\end{array}\right),\\
W_-=\left(\begin{array}{ccc}
i\lambda\theta^-+i\partial_{\theta^-}\phi & -i\lambda\theta^- & -i\sqrt{\lambda}(1+i\theta^-\partial_{\theta^-}\phi) \\
-i\lambda\theta^- & i\lambda\theta^--i\partial_{\theta^-}\phi & i\sqrt{\lambda}(1-i\theta^-\partial_{\theta^-}\phi) \\
-\sqrt{\lambda}(1-i\theta^-\partial_{\theta^-}\phi) & \sqrt{\lambda}(1+i\theta^-\partial_{\theta^-}\phi) & -i\theta^-
\end{array}\right).
\end{array}
\end{equation*}

\subsection{The bosonic Sym--Tafel immersion formula}\label{SecsGBST}
Let us consider the deformation generated by a translation of the spectral parameter $\lambda$, i.e.
\begin{equation}
\beta(\lambda)=1,\qquad F=\Psi^{-1}\partial_{\lambda}\Psi.
\end{equation}
The deformation supermatrices $A_\pm$ and $B_\pm$ take the forms
\begin{equation}
\hspace{-2cm}\begin{array}{l}
\vspace{3mm}A_+=\frac{-1}{4\lambda^{3/2}}\left(\begin{array}{ccc}
0 & 0 & ie^{i\phi} \\
0 & 0 & -ie^{-i\phi} \\
-e^{-i\phi} & e^{i\phi} & 0
\end{array}\right)=\frac{1}{2\lambda}U_+,\qquad A_-=\frac{1}{2\sqrt{\lambda}}\left(\begin{array}{ccc}
0 & 0 & -i \\
0 & 0 & i \\
-1 & 1 & 0
\end{array}\right),\\
\vspace{3mm}B_+=\frac{1}{4\lambda^2}\left(\begin{array}{ccc}
1 & -e^{2i\phi} & i\sqrt{\lambda}D_+\phi e^{i\phi} \\
-e^{-2i\phi} & 1 & i\sqrt{\lambda} D_+\phi e^{-i\phi} \\
-\sqrt{\lambda}D_+\phi e^{i\phi} & -\sqrt{\lambda}D_+\phi e^{i\phi} & 2
\end{array}\right),\\
\vspace{3mm}B_-=\left(\begin{array}{ccc}
1 & -1 & -iD_-\phi/2\sqrt{\lambda} \\
-1 & 1 & -iD_-\phi/2\sqrt{\lambda} \\
-D_-\phi/2\sqrt{\lambda} & -D_-\phi/2\sqrt{\lambda} & 2
\end{array}\right).
\end{array}
\end{equation}
The metric coefficients $g_{ij}$ are given by
\begin{equation}
\begin{array}{l}
g_{11}=g_{22}=g_{13}=g_{31}=g_{24}=g_{42}=g_{33}=g_{44}=0,\\
g_{12}=g_{21}=\frac{1}{4\lambda^2}\left(\cos2\phi-iD_+\phi D_-\phi\cos\phi\right),\\
g_{14}=g_{41}=\frac{1}{4\lambda^2}\left(D_+\phi\sin\phi+i\theta^-\left(\cos2\phi-iD_+\phi\partial_{\theta^-}\phi\cos\phi\right)\right),\\
g_{23}=g_{32}=\frac{1}{4\lambda^2}\left(-D_-\phi\sin\phi+i\theta^+\left(\cos2\phi-i\partial_{\theta^+}\phi D_-\phi\cos\phi\right)\right),\\
g_{34}=-g_{43}=\frac{i}{4\lambda^2}(\cos\phi+\sin\phi\left(\theta^+\partial_{\theta^+}\phi+\theta^-\partial_{\theta^-}\phi\right)\\
\hspace{3cm}+i\theta^+\theta^-\left(\cos2\phi-i\partial_{\theta^+}\phi\partial_{\theta^-}\phi\cos\phi\right))
\end{array}
\end{equation}
and the metric
\begin{equation}
g=\left(\begin{array}{cccc}
0 & g_{12} & 0 & g_{14} \\
g_{12} & 0 & g_{23} & 0 \\
0 & g_{23} & 0 & g_{34} \\
g_{14} & 0 & -g_{34} & 0
\end{array}\right)
\end{equation}
takes its values in the $GL(2\vert2,\mathbb{G})$ Lie supergroup. Hence, $g$ is invertible.

A normal unit vector is given by
\begin{equation}
N=\Psi^{-1}\left(\begin{array}{ccc}
1 & 0 & 0 \\
0 & -1 & 0 \\
0 & 0 & 0
\end{array}\right)\Psi
\end{equation}
and the coefficients of the second fundamental form are given by
\begin{equation}
\begin{array}{l}
b_{11}=b_{12}=b_{21}=b_{22}=b_{13}=b_{31}=b_{24}=b_{42}=b_{33}=b_{44}=0,\\
b_{14}=b_{41}=-iD_+\phi\frac{1}{2\lambda}\cos\phi,\qquad b_{23}=b_{32}=iD_-\phi\frac{1}{2\lambda}\cos\phi,\\
b_{34}=-b_{43}=\frac{1}{2\lambda}\left(\sin\phi-\cos\phi\left(\theta^+\partial_{\theta^+}\phi+\theta^-\partial_{\theta^-}\phi\right)\right)
\end{array}
\end{equation}
such that
\begin{equation}
b=\left(\begin{array}{cccc}
0 & 0 & 0 & b_{14} \\
0 & 0 & b_{23} & 0 \\
0 & b_{23} & 0 b_{34} \\
b_{14} & 0 & -b_{34} & 0
\end{array}\right).
\end{equation}
The mean curvature depends linearly on the spectral parameter $\lambda$ and is given by
\begin{equation*}
\hspace{-1cm}\begin{array}{r}
H=4\lambda\sin\phi\left(-D_+\phi D_-\phi\tan^2\phi\sec2\phi-i\left(\theta^+\partial_{\theta^+}\phi+\theta^-\partial_{\theta^-}\phi\right)\tan\phi\sec\phi\right.\\
\left.+\theta^+\theta^-\cos2\phi\sec^2\phi+i\theta^+\partial_{\theta^+}\phi\theta^-\partial_{\theta^-}\phi\sec\phi(2\tan^2\phi+1)\right)
\end{array}
\end{equation*}
and the Gaussian curvature $K$ is not defined since $b$ is not invertible.

\subsection{The translation in $x_\pm$ symmetry deformation}\label{SecsGBpr}
Let us consider the bosonic vector field
\begin{equation}
\omega_R=\partial_{x_\pm},\qquad F=\Psi^{-1}D_{x_\pm}\Psi
\end{equation}
which generates the translation in the direction of $x_\pm$. This deformation is equivalent to a bosonic gauge deformation given by $S=V_\pm$. The deformation supermatrices $A_\pm$ and $B_\pm$ are given by
\begin{equation}
\hspace{-2.5cm}\begin{array}{l}
\vspace{3mm}A_+=\frac{\partial_{x_\pm}\phi}{2\sqrt{\lambda}}\left(\begin{array}{ccc}
0 & 0 & -e^{i\phi} \\
0 & 0 & -e^{-i\phi} \\
ie^{-i\phi} & ie^{i\phi} & 0
\end{array}\right),\qquad A_-=i\partial_{x_\pm}D_-\phi\left(\begin{array}{ccc}
1 & 0 & 0 \\
0 & -1 & 0 \\
0 & 0 & 0
\end{array}\right),\\
\vspace{3mm}B_+=\frac{1}{2\lambda}\left(\begin{array}{cc}
0 & i\partial_{x_\pm}\phi e^{2i\phi} \\
-i\partial_{x_\pm}\phi e^{-2i\phi} & 0 \\
\sqrt{\lambda}e^{-i\phi}(-i\partial_{x_\pm}\phi D_+\phi+\partial_{x_\pm}D_+\phi) & \sqrt{\lambda}e^{i\phi}(i\partial_{x_\pm}\phi D_+\phi+\partial_{x_\pm}D_+\phi) 
\end{array}\right.\\
\vspace{3mm}\hspace{7cm}\left.\begin{array}{c}
 \sqrt{\lambda}e^{i\phi}(\partial_{x_\pm}\phi D_+\phi-i\partial_{x_\pm}D_+\phi) \\
 -\sqrt{\lambda}e^{-i\phi}(\partial_{x_\pm}\phi D_+\phi+i\partial_{x_\pm}D_+\phi) \\ 
 0
\end{array}\right),\\
B_-=\left(\begin{array}{ccc}
i\partial_{x_\pm}\partial_{x_-}\phi & 0 & -i\sqrt{\lambda}\partial_{x_\pm}D_-\phi \\
0 & -i\partial_{x_\pm}\partial_{x_-}\phi & -i\sqrt{\lambda}\partial_{x_\pm}D_-\phi \\
-\sqrt{\lambda}\partial_{x_\pm}D_-\phi & -\sqrt{\lambda}\partial_{x_\pm}D_-\phi & 0
\end{array}\right).
\end{array}
\end{equation}
The metric coefficients $g_{ij}$ are given by
\begin{equation}
\hspace{-2cm}\begin{array}{l}
g_{33}=g_{44}=0,\qquad g_{11}=\frac{1}{4\lambda^2}(\partial_{x_\pm}\phi)^2,\qquad g_{22}=-(\partial_{x_\pm}\partial_{x_-}\phi)^2,\\
g_{12}=g_{21}=\left(-i\sin\phi\partial_{x_\pm}\phi D_+\phi+i\cos\phi\partial_{x_\pm}D_+\phi\right)\partial_{x_\pm}D_-\phi,\\
g_{13}=g_{31}=\frac{1}{2\lambda}\partial_{x_\pm}\phi D_+\partial_{x_\pm}\phi+\frac{1}{4\lambda^2}i\theta^+(\partial_{x_\pm}\phi)^2,\\
g_{14}=g_{41}=i\theta^-g_{12}=\theta^-\left(\sin\phi\partial_{x_\pm}\phi D_+\phi-\cos\phi\partial_{x_\pm}D_+\phi\right)\partial_{x_\pm}\partial_{\theta^-}\phi,\\
g_{23}=g_{32}=\partial_{x_\pm}\phi\partial_{x_\pm}D_-\phi\cos\phi+\theta^+\left(\sin\phi\partial_{\theta^+}\phi-\cos\phi\partial_{x_\pm}\partial_{\theta^+}\phi\right)\partial_{x_\pm}D_-\phi,\\
g_{24}=g_{42}=-\partial_{x_\pm}D_-\phi\partial_{x_\pm}\partial_{x_-}\phi-i\theta^-(\partial_{x_\pm}\partial_{x_-}\phi)^2\\
g_{34}=-g_{43}=-i\theta^-g_{23}=-i\theta^-\partial_{x_\pm}\phi\partial_{x_\pm}\partial_{\theta^-}\phi\cos\phi\\
\hspace{5cm}+i\theta^+\theta^-\left(\sin\phi\partial_{\theta^+}\phi-\cos\phi\partial_{x_\pm}\partial_{\theta^+}\phi\right)\partial_{x_\pm}\partial_{\theta^-}\phi.
\end{array}
\end{equation}

An $\mathfrak{sl}(2\vert1,\mathbb{G})$-valued bosonic normal unit vector is given by
\begin{equation}
N=\Psi^{-1}\left(\begin{array}{ccc}
i & 0 & 0 \\
0 & i & 0 \\
0 & 0 & 2i
\end{array}\right)\Psi.
\end{equation}
Therefore, the second fundamental form's coefficients are given by
\begin{equation}
\hspace{-2cm}\begin{array}{l}
b_{12}=b_{21}=b_{14}=b_{41}=b_{23}=b_{32}=b_{33}=b_{34}=b_{43}=b_{44}=0,\\
b_{11}=\frac{-1}{2\lambda}\partial_{x_\pm}D_+\phi D_+\phi,\qquad b_{13}=b_{31}=\frac{i}{2\lambda}\left(\partial_{x_\pm}\phi D_+\phi-\theta^+\partial_{x_\pm}\partial_{\theta^+}\phi\partial_{\theta^+}\phi\right),\\
b_{22}=2\lambda\partial_{x_\pm} D_-\phi D_-\phi,\qquad b_{24}=b_{42}=i\theta^-b_{22}=2i\lambda\theta^-\partial_{x_\pm}\partial_{\theta^-}\phi\partial_{\theta^-}\phi,
\end{array}
\end{equation}
such that
\begin{equation}
b=\left(\begin{array}{cccc}
b_{11} & 0 & b_{13} & 0 \\
0 & b_{22} & 0 & i\theta b_{22} \\
b_{13} & 0 & 0 & 0 \\
0 & i\theta^- b_{22} & 0 & 0
\end{array}\right).
\end{equation}
The mean and Gaussian curvatures are not defined since $g$ is not invertible.

\subsection{A bosonic gauge deformation}\label{SecsGBS}
Let us consider the bosonic $\mathfrak{sl}(2\vert1,\mathbb{G})$-valued gauge
\begin{equation}
S=D_-\phi U_-=\sqrt{\lambda}D_-\phi\left(\begin{array}{ccc}
0 & 0 & -i \\
0 & 0 & i \\
-1 & 1 & 0
\end{array}\right)
\end{equation}
with the associated deformation supermatrices
\begin{equation*}
\hspace{-2.5cm}\begin{array}{l}
\vspace{3mm}A_+=\sin\phi\left(\begin{array}{ccc}
D_-\phi & 0 & \sqrt{\lambda} \\
0  & -D_-\phi & -\sqrt{\lambda} \\
-i\sqrt{\lambda} & i\sqrt{\lambda} & 0
\end{array}\right),\qquad A_-=\sqrt{\lambda}\partial_{x_-}\phi\left(\begin{array}{ccc}
0 & 0 & -1 \\
0 & 0 & 1 \\
i & -i & 0 
\end{array}\right),\\
\vspace{3mm}B_+=\left(\begin{array}{cc}
iD_+\phi D_-\phi\cos\phi & 0 \\
0 & -iD_+\phi D_-\phi\cos\phi \\
\frac{-D_-\phi}{4\sqrt{\lambda}}(e^{-2i\phi}-1)-\sqrt{\lambda}\cos\phi D_+\phi & \frac{D_-\phi}{4\sqrt{\lambda}}(e^{2i\phi}-1)+\sqrt{\lambda}\cos\phi D_+\phi
\end{array}\right.\\
\vspace{3mm}\hspace{8cm}\left.\begin{array}{c}
\frac{iD_-\phi}{4\sqrt{\lambda}}(1-e^{2i\phi})+i\sqrt{\lambda}\cos\phi D_+\phi \\
\frac{-iD_-\phi}{4\sqrt{\lambda}}(1-e^{-2i\phi})-i\sqrt{\lambda}\cos\phi D_+\phi \\
0
\end{array}\right),\\
B_-=\sqrt{\lambda}\left(\begin{array}{ccc}
0 & 0 & -i\partial_{x_-}D_-\phi-\partial_{x_-}\phi D_-\phi \\
0 & 0 & i\partial_{x_-}D_-\phi-\partial_{x_-}\phi D_-\phi \\
\partial_{x_-}D_-\phi+i\partial_{x_-}\phi D_-\phi & -\partial_{x_-}D_-\phi+i\partial_{x_-}\phi D_-\phi & 0
\end{array}\right).
\end{array}
\end{equation*}
The metric coefficients $g_{ij}$ take the form
\begin{equation}
\hspace{-2.5cm}\begin{array}{l}
g_{22}=g_{24}=g_{42}=g_{33}=g_{44}=0,\qquad g_{11}=2iD_+\phi D_-\phi\cos\phi\sin^2\phi,\\
g_{12}=g_{21}=4iD_-\phi\partial_{x_-}D_-\phi\sin^2\phi,\\
g_{13}=g_{31}=-\sin^3\phi D_-\phi-2\theta^+\partial_{\theta^+}\phi D_-\phi\cos\phi\sin^2\phi,\\
g_{14}=g_{41}=-\sin^2\phi(\partial_{x_-}\phi D_-\phi+4\theta^-\partial_{\theta^-}\phi\partial_{x_-}\partial_{\theta^-}\phi),\\
g_{23}=g_{32}=i\theta^+g_{12}=-4\theta^+ D_-\phi\partial_{x_-}D_-\phi\sin^2\phi,\\
g_{34}=-g_{43}=i\theta^+g_{14}=-i\theta^+\sin^2\phi\partial_{x_-}\phi D_-\phi-4i\theta^+\theta^-\partial_{\theta^-}\phi\partial_{x_-}\partial_{\theta^-}\phi\sin^2\phi.
\end{array}
\end{equation}
A normal unit vector $N$ is given by the bosonic $\mathfrak{sl}(2\vert1,\mathbb{G})$-valued supermatrix
\begin{equation}
N=\Psi^{-1}\left(\begin{array}{ccc}
i & 0 & 0 \\
0 & i & 0 \\
0 & 0 & 2i
\end{array}\right)\Psi
\end{equation}
and the second fundamental form's coefficients are
\begin{equation}
\hspace{-2cm}\begin{array}{l}
b_{12}=b_{21}=b_{22}=b_{23}=b_{32}=b_{24}=b_{42}=b_{33}=b_{44}=0,\\
b_{11}=\frac{i}{2\lambda}D_+\phi D_-\phi\sin\phi,\qquad b_{13}=b_{31}=i\theta^+b_{11}=\frac{-1}{2\lambda}\theta^+\partial_{\theta^+}\phi D_-\phi\sin\phi,\\
b_{14}=b_{41}=-2\lambda\cos\phi D_+\phi,\qquad b_{34}=-b_{43}=-2i\lambda(\sin\phi-\theta^+\partial_{\theta^+}\phi\cos\phi).
\end{array}
\end{equation}
The curvatures are not defined since $g$ is not an element of the $GL(2\vert2,\mathbb{G})$ Lie supergroup.

\subsection{The supersymmetric transformation deformation}\label{SecsGFpr}
Let us consider the SUSY transformation generators
\begin{equation}
J_\pm=\partial_{\theta^+\pm}+i\theta^\pm\partial_{x_\pm},
\end{equation}
which generate the associated transformations
\begin{equation}
x_\pm\rightarrow x_\pm-i\theta^\pm\underline{\xi}^\pm,\qquad \theta^\pm\rightarrow\theta^\pm+i\underline{\xi}^\pm,
\end{equation}
where $\underline{\xi}^\pm$ are arbitrary fermonic parameters. The differential operators $J_\pm$ satisfy the properties
\begin{equation}
\hspace{-1cm}J_\pm^2=i\partial_{x_\pm},\qquad \lbrace J_+,J_-\rbrace=0,\qquad \lbrace J_\pm,D_\pm\rbrace=0,\qquad \lbrace J_\pm,D_\mp\rbrace=0.
\end{equation}
We can construct a deformed surface $F$ using this common symmetry as
\begin{equation}
F=\Psi^{-1}\mbox{pr}(J_\pm)\Psi.
\end{equation}
The associated deformation supermatrices are given by
\begin{equation*}
\hspace{-2.5cm}\begin{array}{l}
\vspace{3mm}A_+=\frac{\mbox{pr}(J_\pm)\phi}{2\sqrt{\lambda}}\left(\begin{array}{ccc}
0 & 0 & -e^{i\phi} \\
0 & 0 & -e^{-i\phi} \\
ie^{-i\phi} & ie^{i\phi} & 0
\end{array}\right),\qquad A_-=i\mbox{pr}(J_\pm)D_-\phi\left(\begin{array}{ccc}
1 & 0 & 0 \\
0 & -1 & 0 \\
0 & 0 & 0
\end{array}\right),\\
\vspace{3mm}B_+=\frac{-1}{2\sqrt{\lambda}}\left(\begin{array}{cc}
0 & \frac{-i}{\sqrt{\lambda}}\mbox{pr}(J_\pm)\phi e^{2i\phi} \\
\frac{i}{\sqrt{\lambda}}\mbox{pr}(J_\pm)\phi e^{-2i\phi} & 0 \\
(\mbox{pr}(J_\pm)D_+\phi+iD_+\phi\mbox{pr}(J_\pm)\phi)e^{-i\phi} & (\mbox{pr}(J_\pm)D_+\phi-iD_+\phi\mbox{pr}(J_\pm)\phi)e^{i\phi}
\end{array}\right.\\
\vspace{3mm}\hspace{8cm}\left.\begin{array}{c}
(i\mbox{pr}(J_\pm)D_+\phi+D_+\phi\mbox{pr}(J_\pm)\phi)e^{i\phi}\\
(i\mbox{pr}(J_\pm)D_+\phi-D_+\phi\mbox{pr}(J_\pm)\phi)e^{-i\phi}\\
0
\end{array}\right),\\
B_-=\left(\begin{array}{ccc}
i\mbox{pr}(J_\pm)\partial_{x_-}\phi & 0 & -i\sqrt{\lambda}\mbox{pr}(J_\pm)D_-\phi \\
0 & -i\mbox{pr}(J_\pm)\partial_{x_-}\phi & -i\sqrt{\lambda}\mbox{pr}(J_\pm)D_-\phi \\
\sqrt{\lambda}\mbox{pr}(J_\pm)D_-\phi & \sqrt{\lambda}\mbox{pr}(J_\pm)D_-\phi & 0
\end{array}\right).
\end{array}
\end{equation*}
The metric coefficients are
\begin{equation*}
\hspace{-1.5cm}\begin{array}{l}
g_{11}=g_{22}=g_{13}=g_{31}=g_{33}=0,\\
g_{12}=-g_{21}=-i\mbox{pr}(J_\pm)D_+\phi(\cos\phi\mbox{pr}(J_\pm)D_-\phi+\sin\phi D_+\phi\mbox{pr}(J_\pm)\phi),\\
g_{14}=g_{41}=-i\theta^-g_{12}=-\theta^-\mbox{pr}(J_\pm)D_+\phi(\cos\phi\mbox{pr}(J_\pm)D_-\phi+\sin\phi D_+\phi\mbox{pr}(J_\pm)\phi),\\
g_{23}=g_{32}=\mbox{pr}(J_\pm)\phi\mbox{pr}(J_\pm)D_-\phi\cos\phi+\theta^+\mbox{pr}(J_\pm)D_+\phi(\cos\phi\mbox{pr}(J_\pm)D_-\phi\\
\hspace{9cm}+\sin\phi D_+\phi\mbox{pr}(J_\pm)\phi),\\
g_{24}=g_{42}=\mbox{pr}(J_\pm)D_-\phi\mbox{pr}(J_\pm)\partial_{x_-}\phi,\\
g_{44}=-(\mbox{pr}(J_\pm)D_-\phi)^2+2i\theta^-\mbox{pr}(J_\pm)D_-\phi\mbox{pr}(J_\pm)\partial_{x_-}\phi,\\
g_{34}=g_{43}=i\theta^-\mbox{pr}(J_\pm)\phi\mbox{pr}(J_\pm)D_-\phi\cos\phi\\
\hspace{3cm}-i\theta^+\theta^-\mbox{pr}(J_\pm)D_+\phi(\cos\phi\mbox{pr}(J_\pm)D_-\phi+\sin\phi D_+\phi\mbox{pr}(J_\pm)\phi)
\end{array}
\end{equation*}
and $g$ does not take its value in the $GL(2\vert2,\mathbb{G})$ Lie supergroup.

A normal unit vector is given by
\begin{equation}
N=\Psi^{-1}\left(\begin{array}{ccc}
i & 0 & 0 \\
0 & i & 0 \\
0 & 0 & 2i
\end{array}\right)\Psi
\end{equation}
and the second fundamental form coefficients are
\begin{equation}
\begin{array}{l}
b_{12}=b_{21}=b_{14}=b_{41}=b_{23}=b_{32}=b_{33}=b_{34}=b_{43}=b_{44}=0,\\
b_{11}=\frac{-1}{2\lambda}\mbox{pr}(J_\pm)D_+\phi D_+\phi,\qquad b_{22}=2\lambda\mbox{pr}(J_\pm)D_-\phi D_-\phi,\\
b_{13}=b_{31}=\frac{3i}{2\lambda}\mbox{pr}(J_\pm)\phi D_+\phi-\frac{i\theta^+}{2\lambda}\mbox{pr}(J_\pm)D_+\phi D_+\phi,\\
b_{24}=b_{42}=i\theta^-b_{22}=2i\lambda\theta^-\mbox{pr}(J_\pm)D_-\phi D_-\phi,
\end{array}
\end{equation}
such that
\begin{equation}
b=\left(\begin{array}{cccc}
b_{11} & 0 & b_{13} & 0 \\
0 & b_{22} & 0 & i\theta^-b_{22} \\
b_{13} & 0 & 0 & 0 \\
0 & i\theta^-b_{22} & 0 & 0
\end{array}\right).
\end{equation}
The mean curvature is not defined since $g$ is not invertible.

\subsection{A fermionic gauge deformation}
Let us consider the fermionic $\mathfrak{sl}(2\vert1,\mathbb{G})$-valued gauge
\begin{equation}
S=D_-\Phi\left(\begin{array}{ccc}
1 & 0 & 0 \\
0 & 1 & 0 \\
0 & 0 & 2
\end{array}\right),
\end{equation}
which is associated with the deformation supermatrices
\begin{equation}
\begin{array}{l}
\vspace{3mm}A_+=\left(\begin{array}{ccc}
-i\sin\phi & 0 & \frac{-iD_-\phi}{2\sqrt{\lambda}}e^{i\phi} \\
0 & -i\sin\phi & \frac{iD_-\phi}{2\sqrt{\lambda}}e^{-i\phi} \\
\frac{D_-\phi}{2\sqrt{\lambda}}e^{-i\phi} & \frac{-D_-\phi}{2\sqrt{\lambda}}e^{i\phi} & -2i\sin\phi
\end{array}\right),\\
\vspace{3mm}A_-=\left(\begin{array}{ccc}
i\partial_{x_-}\phi & 0 & i\sqrt{\lambda}D_-\phi \\
0 & i\partial_{x_-}\phi & -i\sqrt{\lambda}D_-\phi \\
\sqrt{\lambda}D_-\phi & -\sqrt{\lambda}D_-\phi & 2i\partial_{x_-}\phi
\end{array}\right),\\
\vspace{3mm}B_+=\left(\begin{array}{ccc}
-\sin\phi D_+\phi & 0 & \frac{-i}{2\sqrt{\lambda}}D_+\phi D_-\phi e^{i\phi} \\
0 & -\sin\phi D_+\phi & \frac{-i}{2\sqrt{\lambda}}D_+\phi D_-\phi e^{-i\phi} \\
\frac{D_+\phi D_-\phi}{2\sqrt{\lambda}}e^{-i\phi} & \frac{D_+\phi D_-\phi}{2\sqrt{\lambda}}e^{i\phi} & 2\sin\phi D_+\phi
\end{array}\right),\\
B_-=\partial_{x_-}D_-\phi\left(\begin{array}{ccc}
1 & 0 & 0 \\
0 & 1 & 0 \\
0 & 0 & 2
\end{array}\right).
\end{array}
\end{equation}
The metric coefficients are given by
\begin{equation}
\hspace{-2cm}\begin{array}{l}
g_{11}=g_{22}=0,\qquad g_{12}=-g_{21}=\sin\phi D_+\phi\partial_{x_-}D_-\phi,\\
g_{13}=g_{31}=i\sin^2\phi D_+\phi,\qquad g_{14}=g_{41}=-i\sin\phi D_+\phi(\partial_{x_-}\phi+\theta^-\partial_{x_-}\partial_{\theta^-}\phi),\\
g_{23}=g_{32}=-i\sin\phi\partial_{x_-}D_-\phi(1-\theta^+\partial_{\theta^+}\phi),\qquad g_{24}=g_{42}=i\partial_{x_-}D_-\phi\partial_{x_-}\phi,\\
g_{33}=\sin^2\phi(1-2\theta^+\partial_{\theta^+}\phi),\qquad g_{44}=\partial_{x_-}\phi(\partial_{x_-}\phi-2\theta^-\partial_{x_-}\partial_{\theta^-}\phi),\\
g_{34}=g_{43}=\sin\phi(-\partial_{x_-}\phi+\theta^+\partial_{x_-}\phi\partial_{\theta^+}\phi+\theta^-\partial_{x_-}\partial_{\theta^-}\phi+\theta^+\theta^-\partial_{x_-}\partial_{\theta^-}\phi\partial_{\theta^+}\phi).
\end{array}
\end{equation}
The induced metric on the surface is not invertible since $g_{11}=g_{22}=0$ and $g_{12}$ is nilpotent.

A normal unit vector is given by
\begin{equation}
N=\Psi^{-1}\left(\begin{array}{ccc}
1 & 0 & 0 \\
0 & -1 & 0 \\
0 & 0 & 0
\end{array}\right)\Psi
\end{equation}
and the second fundamental form coefficients are given by
\begin{equation}
\hspace{-1.5cm}\begin{array}{l}
b_{11}=b_{12}=b_{21}=b_{22}=b_{14}=b_{41}=b_{23}=b_{32}=b_{24}=b_{42}=b_{33}=b_{44}=0,\\
b_{13}=b_{31}=\frac{-i}{2\lambda}D_+\phi D_-\phi,\qquad b_{34}=-b_{43}=\frac{1}{2}D_-\phi\sin\phi.
\end{array}
\end{equation}
The mean curvature is not defined since $g$ is not invertible.

\section{Conclusions}\label{SecConc}\setcounter{equation}{0}
In this paper, we have constructed bosonic and fermionic versions of the Fokas--Gel'fand formula for the immersion of $(1+1\vert2)$-supermanifolds in Lie superalgebras. We use three different types of LSPs: the FLSP, the $x_\pm$-LSP and the $\theta^\pm$-LSP, instead of only the FLSP as presented in \cite{BG16}. We provide a link between these three types of LSP so that the $x_\pm$-LSP and $\theta^\pm$-LSP can be computed using only the FLSP. In addition, we investigated the different ZCCs and their relation to the original SUSY integrable system. The use of these three types of LSPs provides a more complete geometric characterization of the supermanifolds under investigation than in the previous paper \cite{BG16}. Using the super Killing form on the four $\mathfrak{g}$-valued tangent vectors $D_jF$, $j=1,2,3,4$, we compute the 16 metric coefficients $g_{ij}$ (instead of the 4 coefficients given in \cite{BG16}). In addition, for a given normal unit vector $N$, we compute the 16 coefficients $b_{ij}$ of the second fundamental form and if the supermatrix $g=[g_{ij}]$ (in equations (\ref{Bg}) or (\ref{Fg})) is invertible, then we can compute the mean curvature $H$ and also the Gaussian curvature $K$ in the bosonic immersion formula.  As an example, we apply these theoretical considerations to the SSGE. As deformations, we consider the translation in the spectral parameter, the translation in the $x_\pm$-direction, the SUSY transformations generated by $J_\pm$ and two gauge deformations, one bosonic and one fermonic. In each case, we obtain a non-trivial geometry. In the paper \cite{BG16}, the authors obtain curve-like metrics in certain cases even if the tangent vectors are linearly independent. In comparable cases, we obtain more complex structures for the metric and the second fundamental form. 

This research can be extended in many directions. One of these directions could be to investigate other SUSY integrable systems, such as the SUSY KdV equation. This particular example would be of great interest since the SUSY KdV equation reduces to the classical KdV equation and the SSGE does not reduce to its classical counterpart unless we impose an additional condition on the bosonic superfield $\phi$. It would also be interesting to investigate if the SUSY FGIF is complete in the sense that there are no additional deformations which can be added to the spectral deformations, the common generalized Lie symmetry deformations and the gauge deformations. Finally, it would be interesting to provide a complete list of geometric classes associated with SUSY integrable systems.

\section*{Acknowledgements}
The author was supported by a doctoral fellowship provided by the Facult\'e des \'Etudes Sup\'erieures et Postdoctorales of the Universit\'e de Montr\'eal and wishes to thank Professor A.~M. Grundland (Centre de Recherches Math\'ematiques, Universit\'e de Montr\'eal) for his useful discussions and support.

\end{document}